\documentclass[12pt,preprint]{aastex}

\shorttitle{Collimation and scattering of the AGN emission in M104}
\shortauthors{Menezes et al.}

\begin{document}

\title{Collimation and scattering of the active galactic nucleus emission in the Sombrero galaxy}

\author{R. B. Menezes, J. E. Steiner, and T. V. Ricci}

\affil{Instituto de Astronomia Geof\'isica e Ci\^encias Atmosf\'ericas, Universidade de S\~ao Paulo, Rua do Mat\~ao 1226, Cidade Universit\'aria, S\~ao Paulo, SP CEP 05508-090, Brazil;}
\email{robertobm@astro.iag.usp.br}

\begin{abstract}

We present an analysis of a data cube of the central region of M104, the Sombrero galaxy, obtained with the GMOS-IFU of the Gemini-South telescope, and report the discovery of collimation and scattering of the active galactic nucleus (AGN) emission in the circumnuclear region of this galaxy. Analysis with PCA Tomography and spectral synthesis revealed the existence of collimation and scattering of the AGN featureless continuum and also of a broad component of the H$\alpha$ emission line. The collimation and scattering of this broad H$\alpha$ component was also revealed by fitting the [NII] $\lambda\lambda6548,6583$ and H$\alpha$ emission lines as a sum of Gaussian functions. The spectral synthesis, together with a $V-I$ image obtained with the \textit{Hubble Space Telescope}, showed the existence of circumnuclear dust, which may cause the light scattering. We also identify a dusty feature that may be interpreted as a torus/disk structure. The existence of two opposite regions with featureless continuum (P.A. $= -18\degr\pm13\degr$ and P.A. $= 162\degr\pm13\degr$) along a direction perpendicular to the torus/disk (P.A. $= 72\degr\pm14\degr$) suggests that this structure is approximately edge-on and collimates the AGN emission. The edge-on torus/disk also hides the broad-line region. The proposed scenario is compatible with the Unified Model and explains why only a weak broad component of the H$\alpha$ emission line is visible and also why many previous studies detected no broad H$\alpha$. The technique used here proved to be an efficient method not only for detecting scattered light but also for testing the Unified Model in low luminosity AGNs.

\end{abstract}

\keywords{galaxies: active --- galaxies: individual(M104) --- galaxies: nuclei --- techniques: spectroscopic}

\section{Introduction}

Low-ionization nuclear emission-line regions (LINERs) are located in the central region of galaxies and show low-ionization emission-line spectra that cannot be attributed exclusively to stars. This class of objects was first identified by \citet{hec80}.	

M104, the Sombrero galaxy, is a highly inclined (nearly edge-on) SA(s)a galaxy at a distance of about 9.2 Mpc. Based on its optical emission-line spectrum, the nucleus of this object has been classified as a LINER \citep{hec80}. Several studies of this galaxy in the optical spectral region have been published, using both ground-based telescopes \citep{ho95,ho97} and the \textit{Hubble Space Telescope} (\textit{HST}; Kormendy et al. 1996). \citet{kor96} found evidence of a broad component in the H$\alpha$ emission line. However, \citet{ho97} modeled the emission lines as sums of Gaussian components and detected no broad H$\alpha$.

Observations have shown that this galaxy has a compact nuclear source ($\le1$ pc) in both the radio \citep{hum84,baj88} and X-ray \citep{fab97,pel02} spectral regions. That is compatible with the existence of an AGN in this galaxy. \citet{ben06}, using mid- and far-infrared data of M104, concluded that the spectral energy distribution (SED) of the nucleus reveals the presence of hot dust. \citet{nic98} analyzed optical and X-ray data of M104 and concluded that the model of photoionization by a low-luminosity AGN \citep{fer83,hal83} is the most adequate to explain the emission-line spectrum of this galaxy.

In this Letter, we analyze a data cube of the nuclear region of M104, obtained with the Integral Field Unit (IFU) of the Gemini Multi-Object Spectrograph (GMOS) of the Gemini-South telescope, and report the discovery of collimation and scattering of the AGN emission in this galaxy.

\section{Observations, reduction and data treatment}

The observations of M104 were made on 2011 February 3. We used the IFU of the GMOS of the Gemini-South telescope, in the one-slit mode. In this mode of observation, the science field of view (FOV) has $5\arcsec \times 3\arcsec\!\!.5$, while the sky FOV (at a distance of $1\arcmin$ from the science FOV) has $5\arcsec \times 1\arcsec\!\!.75$. The final products from observations made with this instrument are data cubes, containing two spatial dimensions and one spectral dimension. Three 10 minute exposures of the nuclear region of M104 were made, with the grating B600-G5323, in a central wavelength of $5920\AA$. The final spectra had a coverage of $4470\AA - 7340\AA$ and a resolution of $R \sim 2600$. The estimated seeing was $0\arcsec\!\!.75$.

The data reduction was made in IRAF environment. At the end of the process, three data cubes were obtained, with spatial pixels (spaxels) of $0\arcsec\!\!.05 \times 0\arcsec\!\!.05$. No sky subtraction was applied because the sky FOV was contaminated with stellar emission from the galaxy.

After the data reduction, we performed a data treatment procedure. First, a correction of the differential atmospheric refraction was applied to all data cubes. In order to combine the three corrected data cubes into one, a median of these data cubes was calculated. After that, a Butterworth spatial filtering \citep{gon02} was applied to all the images of the resulting data cube to remove spatial high-frequency noise. 

We detected the presence of a probable instrumental fingerprint, which appeared in the data cube as large vertical stripes in the images and had a low spectral frequency signature. In order to remove it, we applied the PCA Tomography technique \citep{ste09}. PCA transforms the data originally expressed in correlated coordinates into a new system of uncorrelated coordinates (eigenvectors) ordered by principal components of decreasing variance. PCA Tomography consists in applying PCA to data cubes. In this case, the variables correspond to the spectral pixels and the observables correspond to the spaxels of the data cube. Since the eigenvectors are obtained as a function of the wavelength, they have a shape similar to spectra and, therefore, we call them eigenspectra. On the other hand, since the observables are spaxels, their projections on the eigenvectors are images, which we call tomograms.  The simultaneous analysis of eigenspectra and tomograms allows one to obtain information that, otherwise, would possibly be harder to detect. Using PCA Tomography, we were able to identify and remove the instrumental fingerprint of the data cube analyzed here (J. E. Steiner et al. 2013, in preparation).

\begin{figure}
\plotone{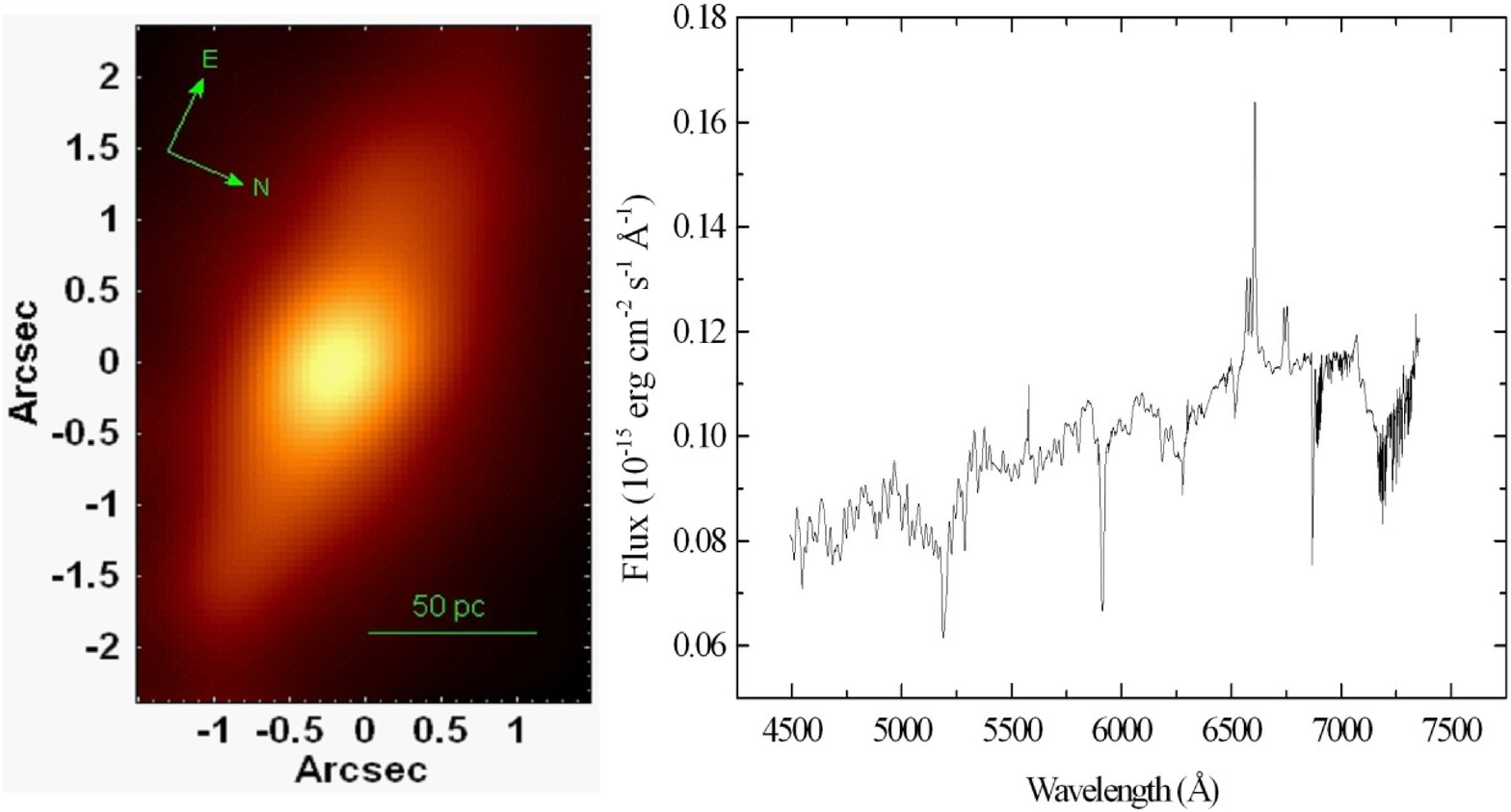}
\caption{Left: image of the final data cube of M104 collapsed along the spectral axis. Right: average spectrum of the final data cube of M104.\label{fig1}}
\end{figure}

Finally, a Richardson-Lucy deconvolution \citep{ric72,luc74} was applied to all the images of the data cube, using a synthetic Gaussian point-spread function (PSF). The PSF of the final data cube has a FWHM of $\sim 0\arcsec\!\!.66$. Figure~\ref{fig1} shows an image of the final data cube of M104 (obtained after the data treatment) collapsed along the spectral axis and also an average spectrum of this data cube.

\section{Data analysis and results}

After the data treatment, we applied again PCA Tomography to the data cube of M104. Figure~\ref{fig2} shows the most significant eigenvectors obtained with this procedure.

Figure 2 shows that eigenspectrum E1 and the corresponding tomogram are very similar, respectively, to the average spectrum and to the image of the data cube of M104 collapsed along the spectral axis. This is expected because eigenvector E1 explains most of the data variance ($99.8356 \%$) and, therefore, reveals the redundancy existent in the data. The tomogram associated with E1 and the sum of all images of the data cube reveal an elongated structure with a position angle (PA) of $8\degr\pm6\degr$. This structure represents the inner stellar disk of this galaxy, which has already been analyzed in previous studies \citep{bur86,ems94, ems96}.

In eigenspectrum E2, we can see correlations to wavelengths corresponding to the main emission lines of the spectrum. We can even see correlations to wavelengths associated with a possible broad component of the H$\alpha$ emission line (with FWZI $\approx 5120$ km s$^{-1}$), with a red wing stronger than the blue one. These characteristics indicate that this eigenvector is correlated to the emission from this galaxy's AGN (in this case a type 1 AGN; Khachikian \& Weedman 1974). In that case, the position of the AGN corresponds to the bright area of the tomogram (correlated to the eigenvector). Eigenspectrum E2 also shows a correlation to the red region and an anti-correlation to the blue region of the spectrum. Such behavior indicates that the bright areas of the tomogram have redder spectra than the dark ones (anti-correlated to the eigenvector), due perhaps to the presence of dust clouds in the regions corresponding to the bright areas. Eigenvector E2 explains $0.7939 \%$ of the data variance.

Eigenspectrum E3 shows correlations to wavelengths corresponding to the blue wings of the main emission lines (and to the red wings of the main absorption lines) and anti-correlations to wavelengths corresponding to the red wings of these lines (and to the blue wings of the absorption lines). This, together with the morphology of the corresponding tomogram, seems to indicate the existence of a possible rotating stellar/gaseous disk around the nucleus of M104; however, some distortions, mainly in the peripheral areas of the image, may indicate the presence of more than one kinematical phenomenon (like a gas outflow, for example). Eigenvector E3 explains $0.1499 \%$ of the data variance.

Eigenspectrum E4 shows a correlation to the blue region of the spectrum and an anti-correlation to the red region of the spectrum. Besides that, we can also see correlations to wavelengths corresponding to the broad component of the H$\alpha$ emission line. These characteristics suggest that the bright areas of the tomogram show bluer spectra than the dark ones and also have a broad H$\alpha$ emission line (although some of these bright areas may show only one of these spectral features and not necessarily both). Therefore, we can see that the bright areas of this tomogram show a typical AGN emission, with the bluer continuum probably representing its featureless continuum. However, none of the bright areas is located at the position of the AGN (revealed by eigenvector E2). Actually, the bright areas in the tomogram associated with E4 are located near the AGN, but above and below the disk. A possible explanation is that the phenomenon revealed by eigenvector E4 corresponds to the emission from the AGN (likely collimated by a torus/disk structure) scattered by dust, molecules, or electrons. Such result has never been found in previous studies of this object. Eigenspectrum E4 is also anti-correlated to a considerably narrow Na I absorption doublet, which is probably related to interstellar absorption. In this case, the dark areas in the corresponding tomogram are associated with interstellar absorption along the inner stellar disk of M104. Eigenvector E4 explains $0.0507 \%$ of the data variance.

Determining the number of significant eigenvectors obtained with PCA Tomography of a data cube may be somewhat subjective. In order to do that, we apply the ``scree test'' \citep{ste09}, which consists of a graph with the variance fractions explained by each eigenvector as a function of the numbers of the corresponding eigenvectors. Figure~\ref{fig3} shows the ``scree test'' obtained with PCA Tomography of the data cube of M104. We can see that the variances explained by the eigenvectors decay considerably until approximately eigenvector E11. From that eigenvector on, the rate of decay decreases significantly and remains nearly constant. This behavior indicates that eigenvectors with order higher or equal to 11 represent mainly noise and, therefore, are not relevant to this analysis. However, among the eigenvectors with order lower than 11, only the ones shown in Figure~\ref{fig2} have structures (peaks and valleys) sufficiently more intense than the surrounding ``noise'', which allowed a more reliable analysis.

After the analysis with PCA Tomography, we applied a spectral synthesis to the spectrum of each spaxel of the data cube, using the Starlight software \citep{cid05}, which fits the stellar spectrum of a given object with a combination of template stellar spectra from a pre-established base. In this work, we used the base of stellar spectra Medium resolution INT Library of Empirical Spectra (MILES; S\'anchez-Bl\'azquez et al. 2006), which has a spectral resolution (FWHM $= 2.3\AA$) very similar to our resolution (FWHM $= 2.1\AA$). Since M104 has an AGN, we added a simple power law ($L_{\nu} \propto \nu^{-\alpha}$, with a spectral index of $\alpha = 1.5$), representing a typical AGN featureless continuum, to the base of stellar spectra. The spectral synthesis provided the values of several parameters, but, due to the purposes of this Letter, only two of them will be discussed here: the interstellar extinction ($A_V$) at the observed object and the flux fraction corresponding to the AGN featureless continuum. 

Before performing the spectral synthesis of the data cube of M104, we took the following steps to prepare the spectra: correction of the interstellar extinction due to the Galaxy, using $A_V = 0.17$ and the reddening law of \citet{car89}; shift to the rest-frame, using $z = 0.003416$; spectral re-sampling with $\Delta\lambda = 1\AA$. We then applied the spectral synthesis to the data cube and elaborated maps of the resulting parameters. The non-subtraction of the sky field during the data reduction had no observable effect (like the detection of a stellar population, approximately constant along the FOV, with solar metallicity) in the results. This is expected, as the observations were made during the new moon. Figure~\ref{fig4} shows maps of the flux associated with the AGN featureless continuum and of $A_V$. In the same figure, we show a $V-I$ image, corresponding to the same FOV of the data cube of M104, obtained with WFPC2 of the \textit{HST}.

Figure~\ref{fig4} shows the presence of a featureless continuum in the FOV of the GMOS-IFU, but only in two regions (P.A. $= -18\degr\pm13\degr$ and P.A. $= 162\degr\pm13\degr$) at distances of $1\arcsec - 2\arcsec$ from the AGN, not at the position of the AGN. In these two regions, the featureless continuum corresponds to approximately $5 \%$ of the total flux. This may be a consequence of a collimation of the AGN featureless continuum, followed by a scattering caused by dust, molecules, or electrons. The map of $A_V$ in Figure~\ref{fig4} reveals the existence of a dust lane (approximately at the east-west direction). This is confirmed by the $V-I$ image \citep{pog00}, obtained with WFPC2 of the \textit{HST}, in Figure~\ref{fig4}. There are, however, a few differences between the map of $A_V$ and the $V-I$ image, which may have been caused by the presence of line-emitting regions (which can affect the $V-I$ image) or areas with lower signal-to-noise ratio (S/N; which can have affected the fitting process, resulting in imprecise values for $A_V$). We can also see a bright nuclear stellar source in the $V-I$ image, which can certainly be interpreted as the AGN featureless continuum. This feature was not detected in the map obtained with the spectral synthesis probably due to the extinction, possibly caused by dust, associated with the poorer spatial resolution of our observation.

After that, we subtracted synthetic stellar spectra (provided by the spectral synthesis) from the spectrum of each spaxel of the data cube. We then extracted spectra from five circular regions (with a radius of $0\arcsec\!\!.15$) of this residual data cube and fitted the emission lines [NII] $\lambda\lambda6548,6583$ and H$\alpha$ in these spectra with a sum of Gaussian functions. Each emission line was fitted by two narrow Gaussians (with different widths). Besides that, we also included one broad Gaussian in the fit, in order to reproduce the possible broad component of H$\alpha$. We assumed that the Gaussians used to fit the [N II] $\lambda6548$ line had the same velocities and 0.328 of the intensities (as established theoretically) of the corresponding Gaussians used to fit the [N II] $\lambda6583$ line. The results are shown in Figure~\ref{fig5}.

Figure~\ref{fig5} reveals that a broad component of H$\alpha$ can be detected in regions 1, 2 and 3; however, no traces of such broad component were detected in regions 4 and 5.

\section{Discussion and conclusions}

\begin{figure}
\epsscale{0.51}
\plotone{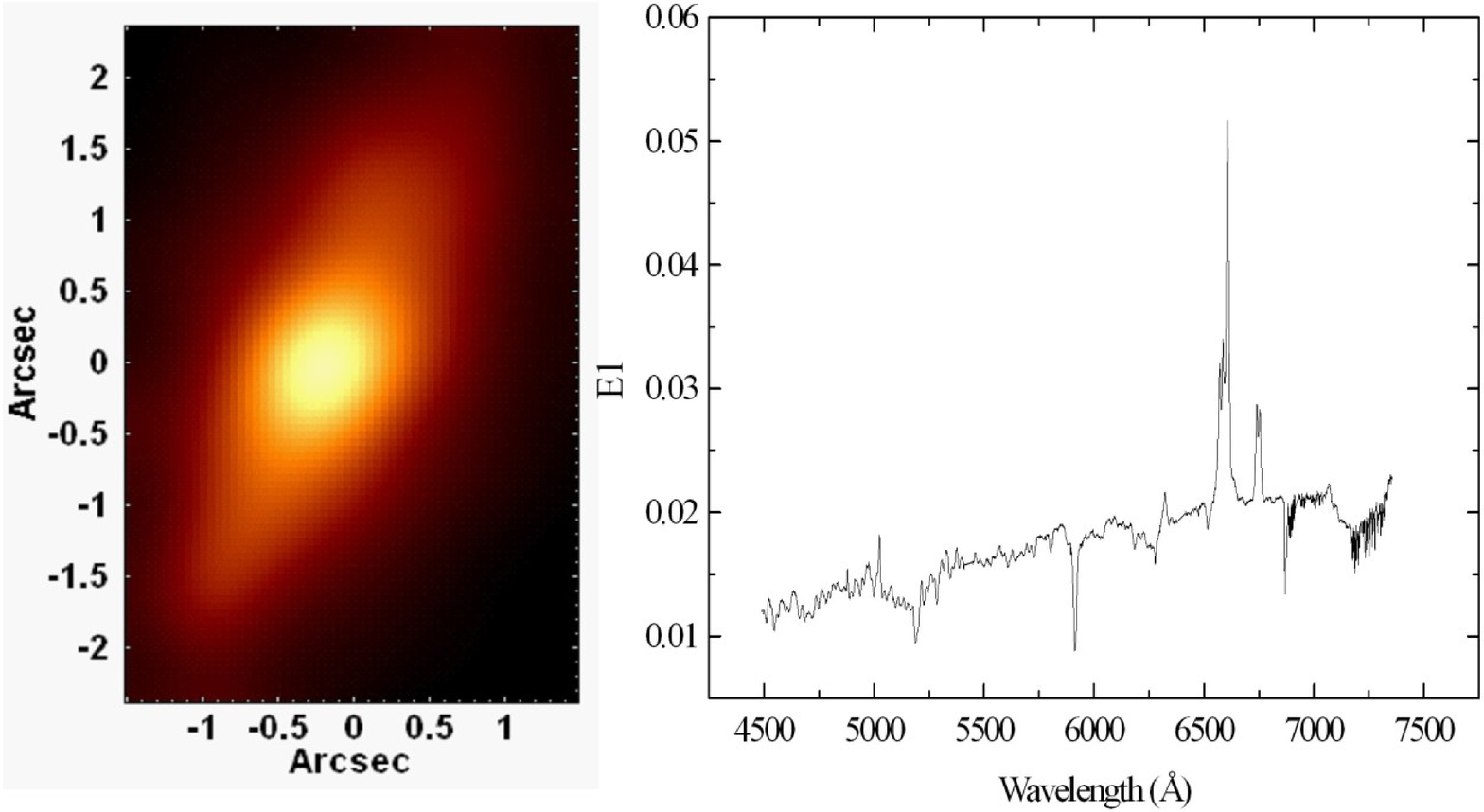}
\plotone{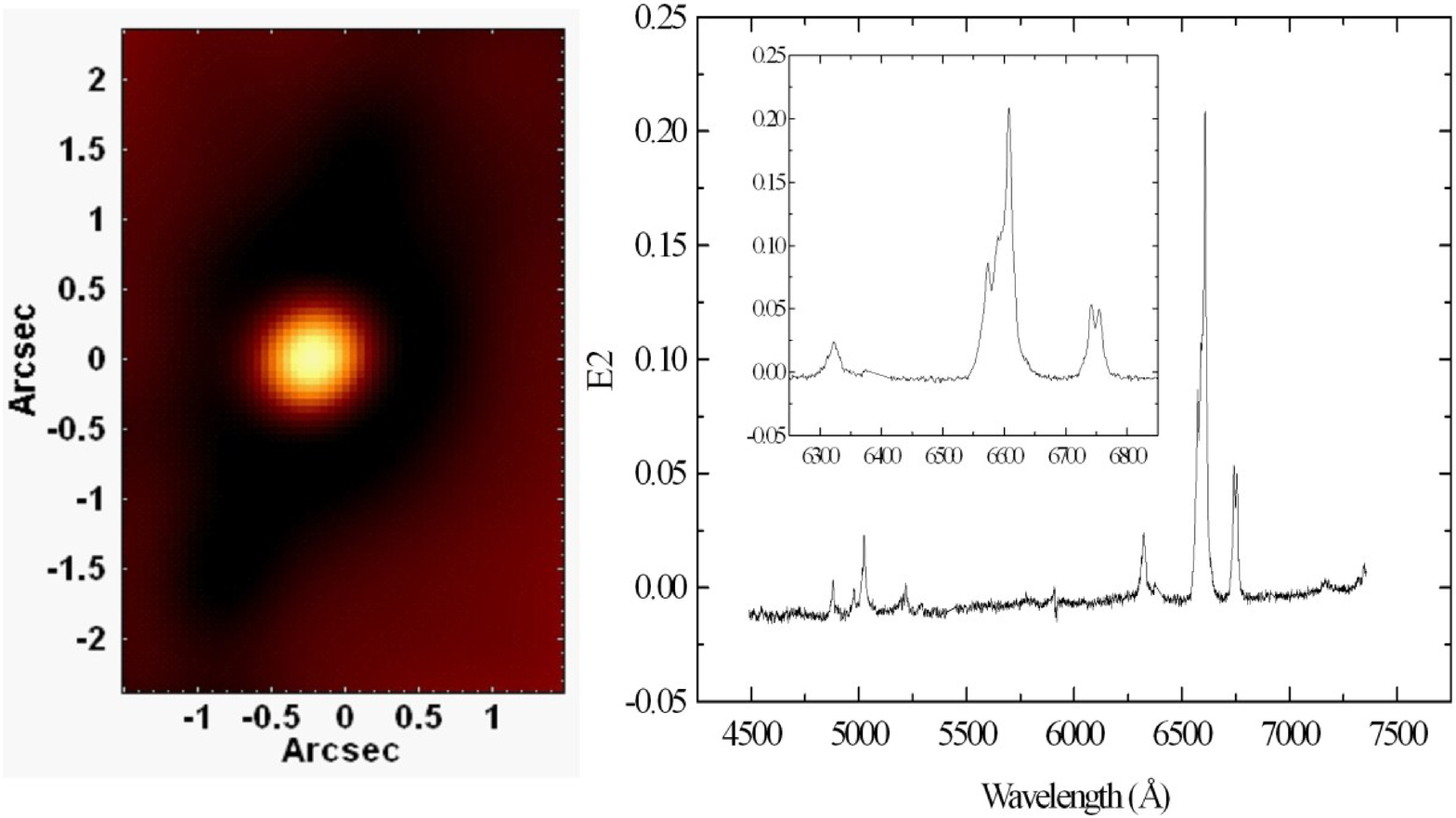}
\plotone{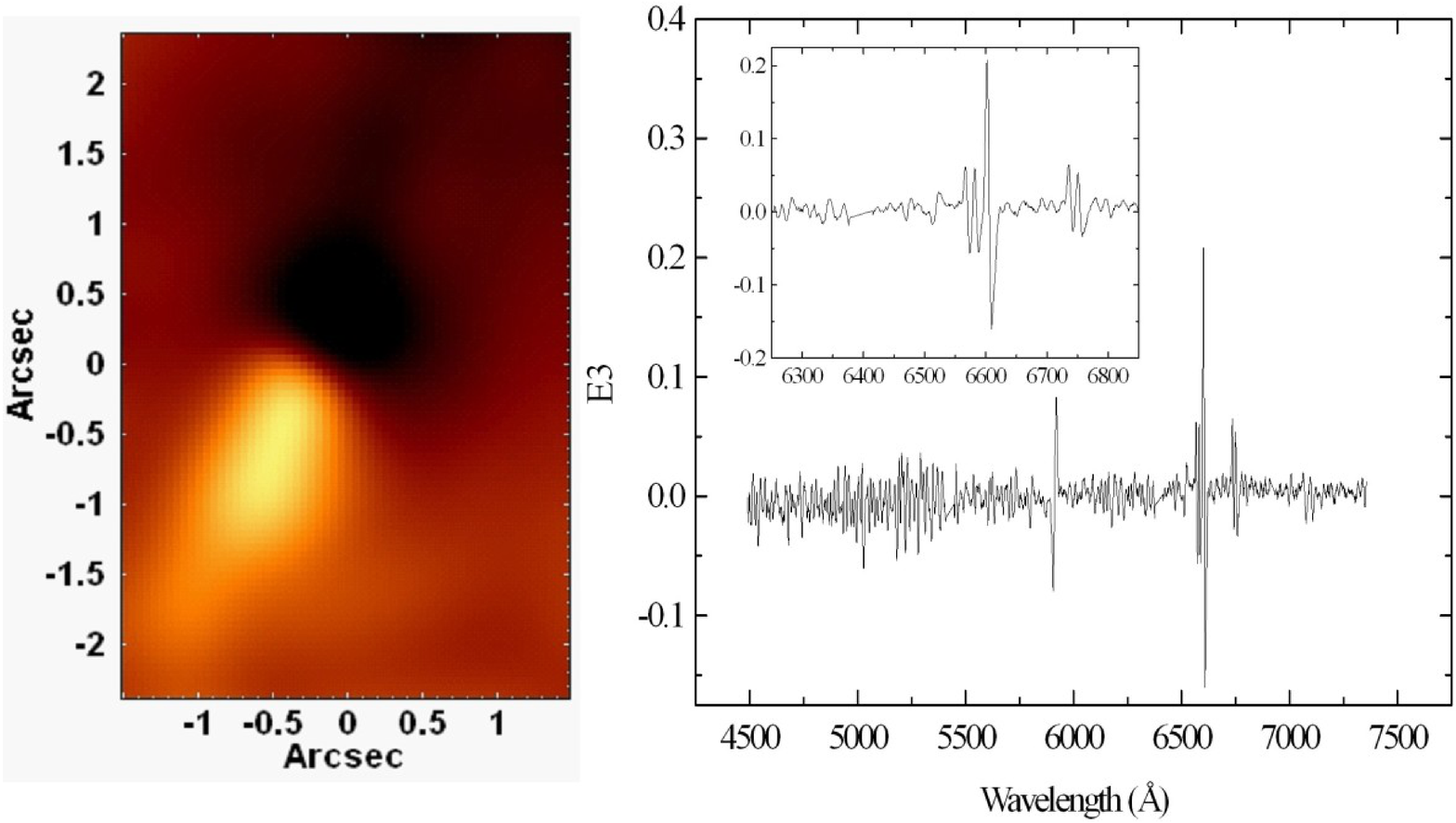}
\plotone{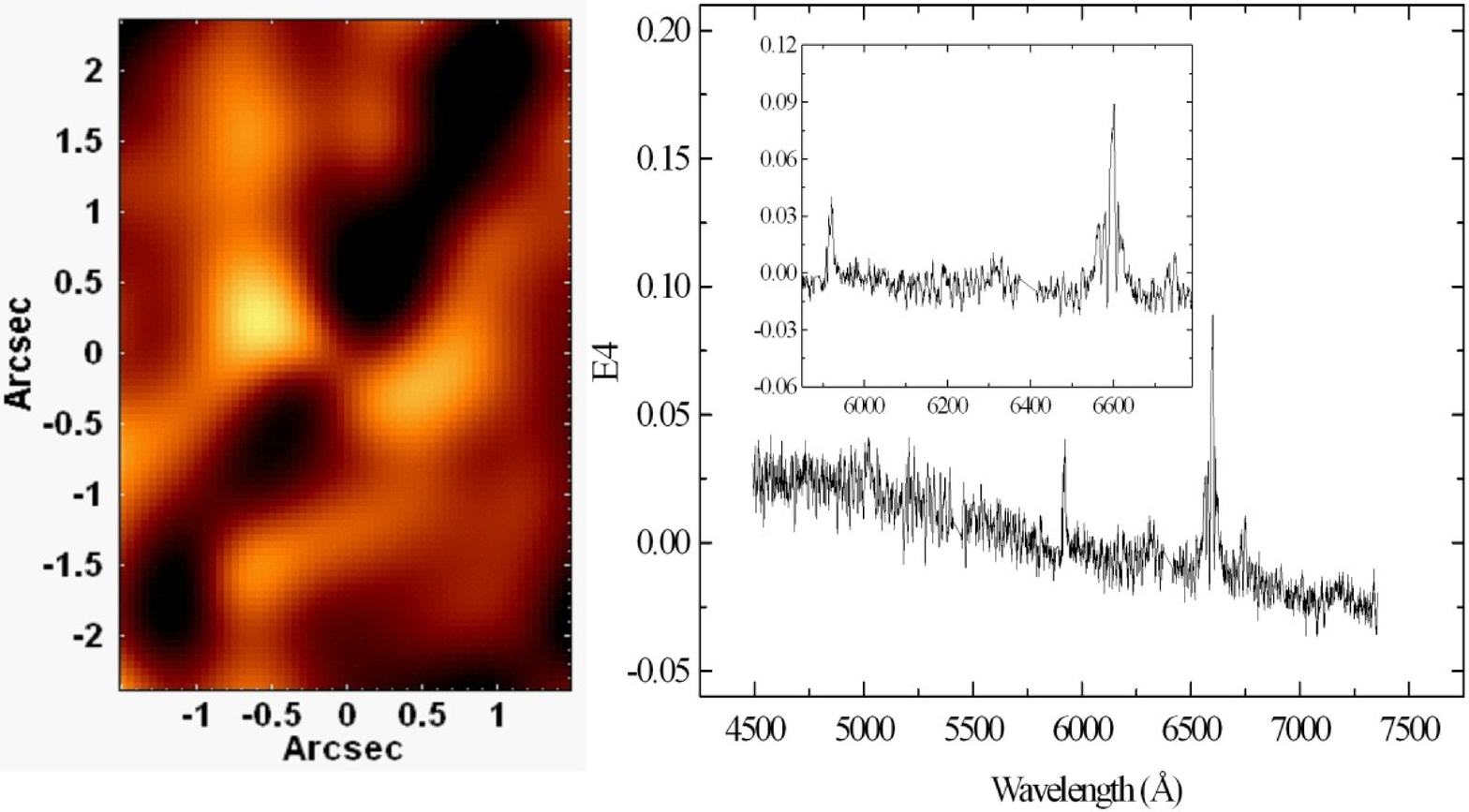}
\caption{Eigenspectra E1, E2, E3, and E4 and the corresponding tomograms, obtained with PCA Tomography. Magnifications of eigenspectra E2, E3, and E4 are also shown.\label{fig2}}
\end{figure}

\begin{figure}
\epsscale{1.0}
\plotone{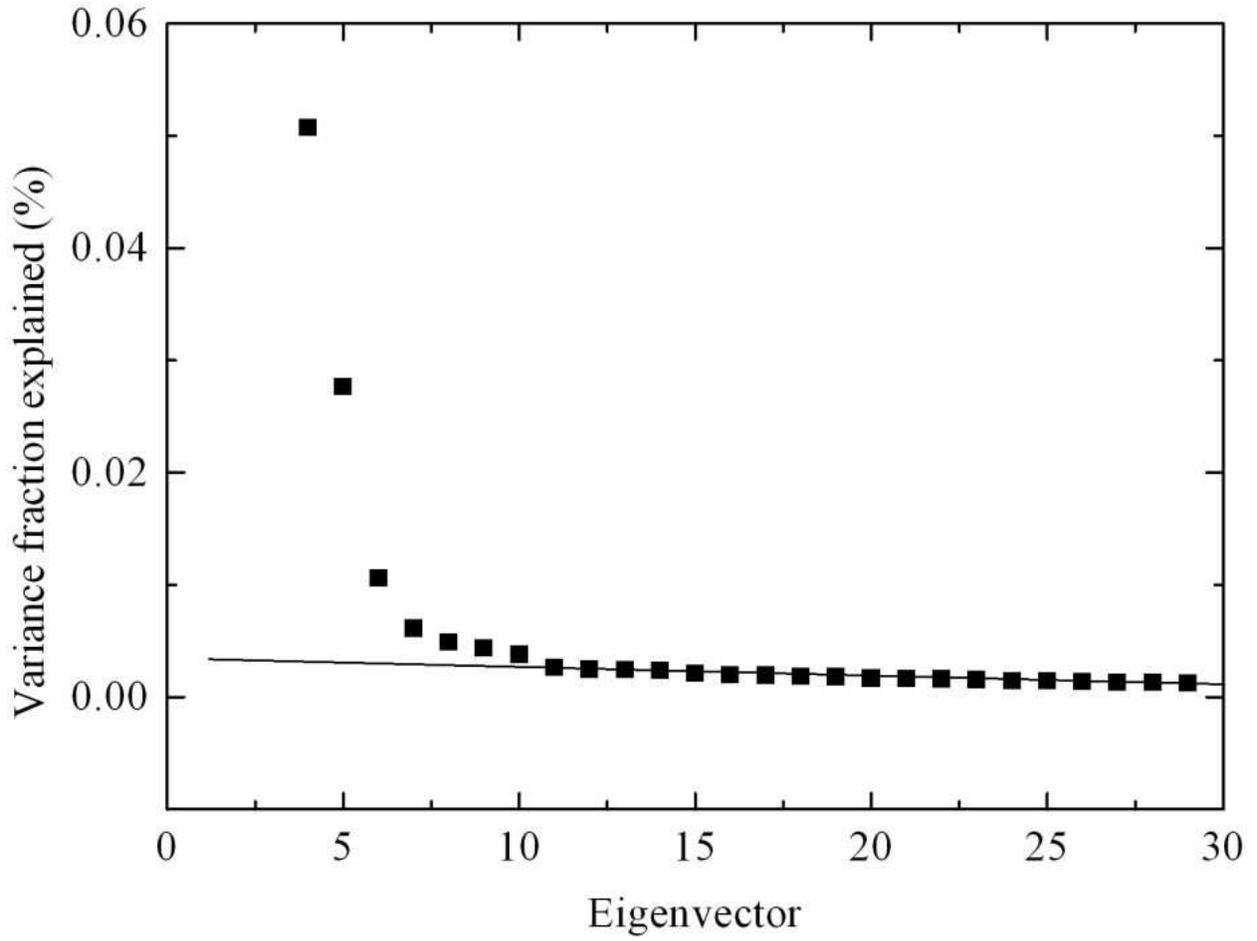}
\caption{``Scree test'' of the eigenvectors obtained with PCA Tomography.\label{fig3}}
\end{figure}

\begin{figure}
\plotone{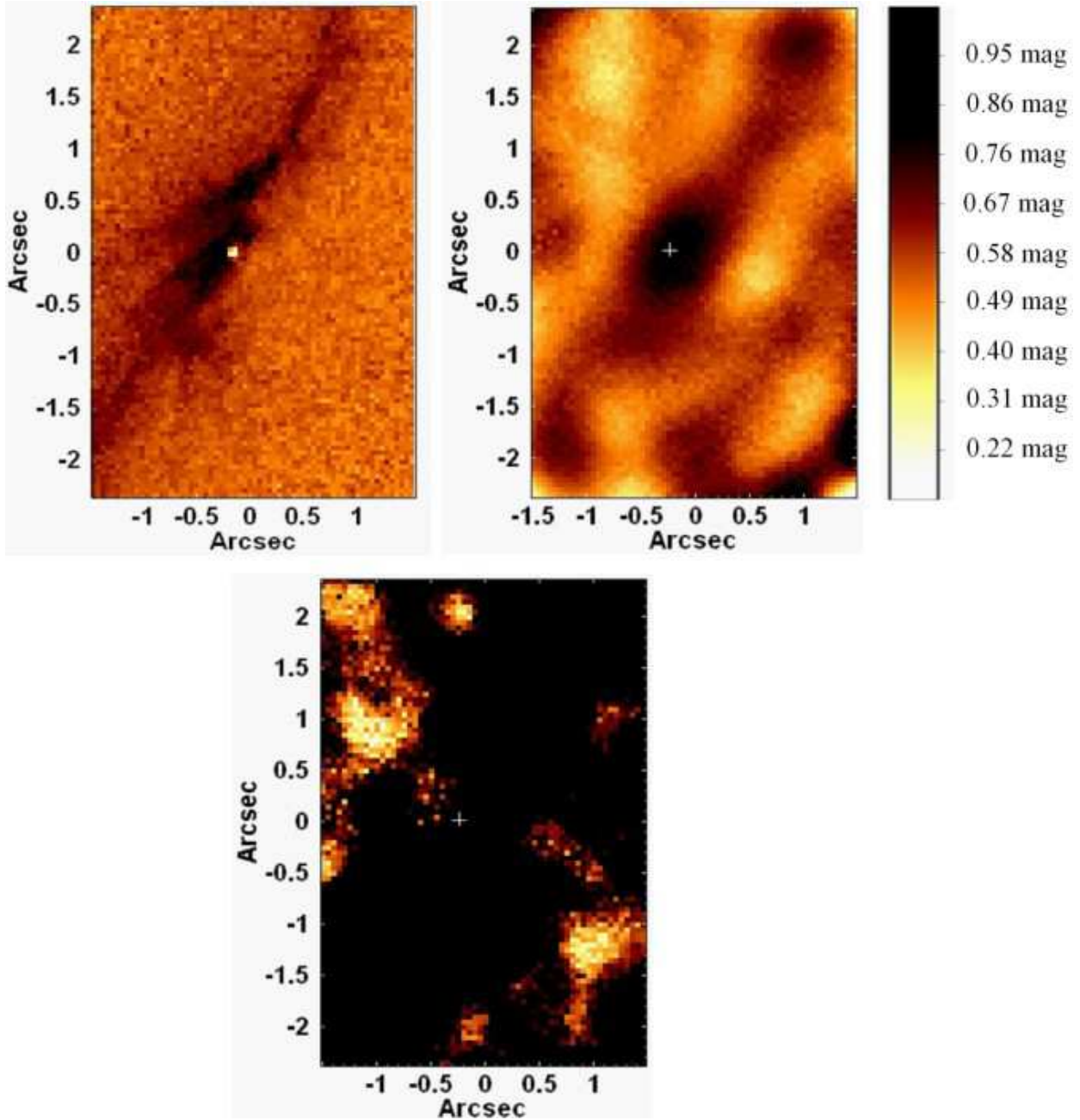}
\caption{Left above: $V-I$ image, obtained with WFPC2 of the \textit{HST}, with the same FOV of the data cube of M104. Darker areas denote regions with dust extinction. Right above: map of $A_V$, obtained with the spectral synthesis. Bottom: map of the flux associated with the AGN featureless continuum, obtained with the spectral synthesis. The position of the AGN (from tomogram E2) is indicated with a $+$ sign.\label{fig4}}
\end{figure}

\begin{figure}
\epsscale{0.78}
\plotone{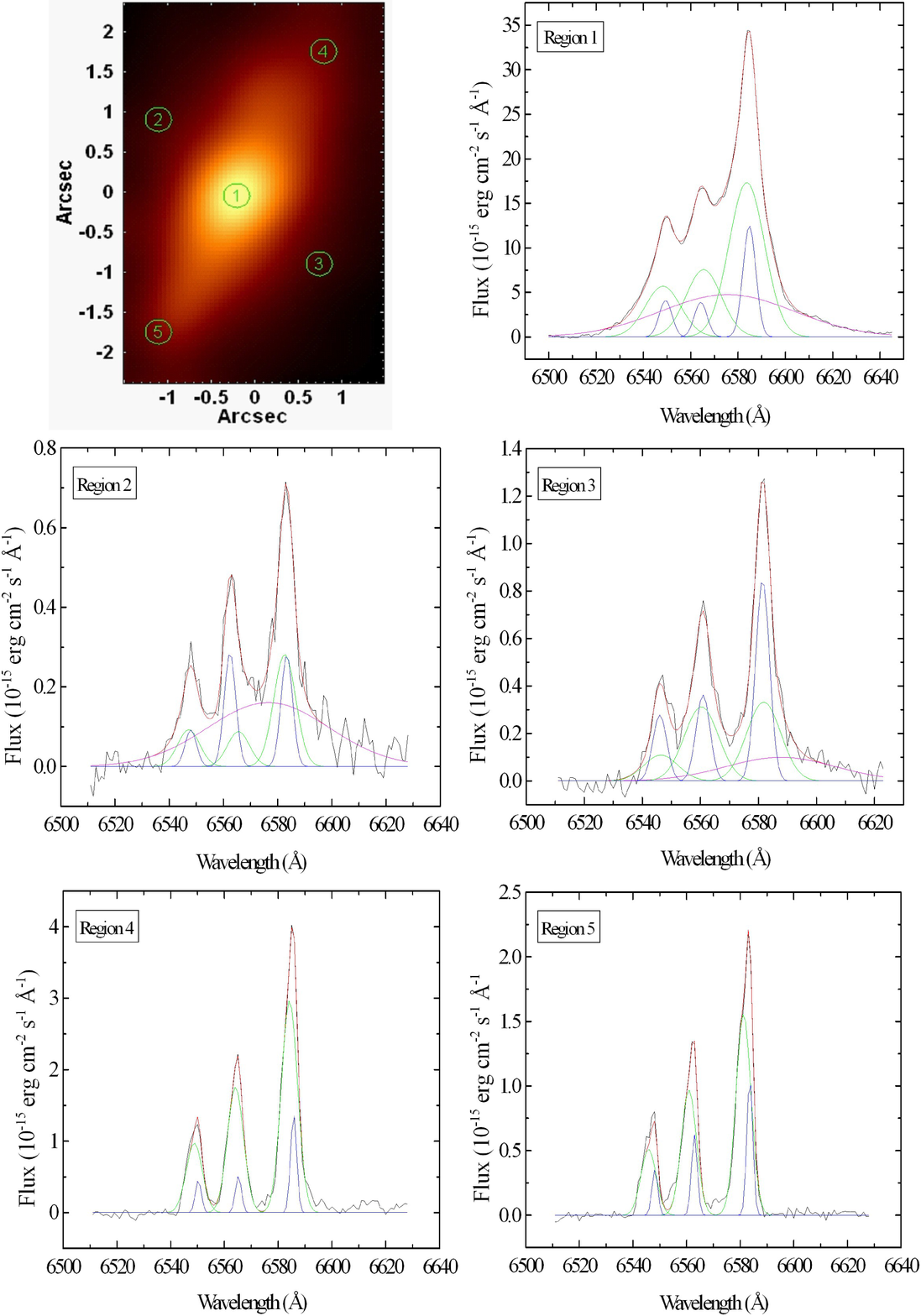}
\caption{Gaussian fits applied to the emission lines [N II] $\lambda\lambda6548,6583$ and H$\alpha$ of the spectra extracted from five circular regions in the data cube of M104 without the stellar continuum. The Gaussians in blue and green represent the narrow components of the emission lines. The Gaussian in magenta represents the broad component of H$\alpha$. The final fits are shown in red.\label{fig5}}
\end{figure}

We can see that the morphologies of the tomogram associated with eigenvector E4 (Figure~\ref{fig2}) and of the map of the flux associated with the AGN featureless continuum (Figure~\ref{fig4}) seem to be compatible. This strongly suggests that the two images reveal the same phenomenon: collimation followed by scattering of the AGN emission. Since eigenvector E4 shows an apparent broad component of the H$\alpha$ emission line, we conclude that, besides the AGN featureless continuum, a possible emission from the broad-line region (BLR) of the AGN is also collimated and scattered. This hypothesis is confirmed by the results shown in Figure~\ref{fig5}, as the broad component of the H$\alpha$ line was only detected at the position of the AGN, south and north from the stellar disk (regions 1, 2, and 3, respectively), but not along the disk (regions 4 and 5). However, some important aspects must be considered here. First of all, the collimation of the AGN emission is more clearly seen in the map of the AGN featureless continuum than in the tomogram associated with eigenvector E4. A possible explanation is that eigenvector E4 may be subject to ``contaminations'' by other phenomena, which could affect the visualization of the collimated emission from the AGN in the corresponding tomogram. Another important point is that the map of the AGN featureless continuum shows emission only in regions $1\arcsec - 2\arcsec$ from the AGN, but the tomogram associated with E4 indicates the existence of emission in regions much closer to the AGN. This may be due to the fact that, unlike the map in Figure~\ref{fig4}, eigenvector E4 reveals not just the AGN featureless continuum, but also the broad component of H$\alpha$, which is less affected by dust extinction and, therefore, can be detected closer to the AGN. 

The detection of a broad component of H$\alpha$ suggests that M104 should be classified as a LINER 1. The FWZI we obtained for this broad H$\alpha$ (FWZI $\approx 5120$ km s$^{-1}$) is very similar to the value found by \citet{kor96} (FWZI $\approx 5200$ km s$^{-1}$). Our result, however, is not compatible with that obtained by \citet{ho97} who, using ground-based observations (with a spatial resolution considerably worse than ours), found no evidence of a broad component of H$\alpha$. 

Although our spatial resolution is not as high as the resolution of the \textit{HST}, the morphologies of the $V-I$ image and of the $A_V$ map in Figure~\ref{fig4} appear to be compatible. The $V-I$ image shows that most of the dust in the vicinity of the AGN in M104 is concentrated in a dense structure (P.A. $= 72\degr\pm14\degr$), with an outer radius of $0\arcsec\!\!.40$. This corresponds to $r = 18$ pc, well within the radius of influence of the black hole, which is $\sim 55$ pc for a $5 \times 10^8 M_{\sun}$ black hole mass \citep{jar11}. This structure may be interpreted as an obscuring torus/disk around the BLR. The fact that this structure is perpendicular to the line connecting the two regions with featureless continuum (P.A. $= -18\degr\pm13\degr$ and P.A. $= 162\degr\pm13\degr$) suggests not only that the torus/disk collimates the AGN emission, but also that it should be approximately edge-on. Figure 1(d) of \citet{pog00} shows that the amount of dust decreases considerably at farther regions from the nucleus (almost no dust is detected at $10\arcsec$ from the nucleus), indicating that the hypothesis of this dust be foreground is very unlikely.

Three mechanisms are capable of producing the observed scattering of the AGN emission. The first is the Thompson scattering by free electrons. Eigenvectors E1 and E3 show that there is diffuse emission of ionized gas around the AGN of this galaxy. This indicates the presence of free electrons, which could produce Thompson scattering at some level. The second mechanism is the Rayleigh scattering by molecules. Using AO-assisted IFU spectroscopy in the $K$ band, we have detected a weak molecular emission of H$_2$ in the nuclear region of M104 (Menezes 2012; R. B. Menezes \& J. E. Steiner 2013, in preparation). This reveals the presence of molecules, which could be responsible, at least partially, for the scattering of the AGN emission. However, the presence of dust around the AGN of this galaxy points toward the third mechanism to be considered: scattering by dust. Indeed, the map of $A_V$ and the $V-I$ image reveal plenty of evidence of dust in the circum-nuclear region of M104, making the scattering of the AGN emission by dust the most likely scenario here. The scattered light has a blue continuum very likely because it is intrinsically blue and not necessarily because of the scattering process. We predict that this scattered light should be polarized, as observed in other LINERs \citep{bar99}.

Considering the way the featureless continuum and the broad H$\alpha$ are collimated and scattered, as well as the apparent distribution of dust in the central region of this galaxy, we conclude that the scenario we propose here is self-consistent. This scenario is compatible with the Unified Model \citep{ant93, urr95}; furthermore, the existence of a nearly edge-on dust torus/disk structure around the BLR explains why only a weak broad component of the H$\alpha$ emission line is visible and also why some studies \citep{ho97} detected no broad H$\alpha$. PCA Tomography (which has already been used by Ricci et al. 2011 to detect scattered light from a low luminosity AGN), combined with spectral synthesis, proved to be an efficient method not only for detecting scattered light but also for testing the Unified Model in low luminosity AGNs.

\acknowledgments

This work is based on observations obtained at the Gemini Observatory. We thank FAPESP for support under grants 2012/02268-8 (R.B.M.) and 2008/06988-0 (T.V.R) and also an anonymous referee for valuable comments about this Letter.

{\it Facilities:} \facility{Gemini:Gillett(GMOS)}.

\clearpage

\clearpage

\clearpage

\clearpage

\clearpage

\end{document}